\documentclass[conference, 10pt]{IEEEtran}
\usepackage[T1]{fontenc}
\usepackage[utf8]{inputenc}
\usepackage{bm}
\usepackage[cmex10]{amsmath}
\usepackage{amssymb,amsthm}
\usepackage{mathtools}
\usepackage{hyperref}
\usepackage{color}

\usepackage{graphicx}
\graphicspath{{figures/}}

\usepackage[backend=biber,style=ieee,natbib=false,mincrossrefs=1000]{biblatex}
\AtBeginDocument{\toggletrue{blx@useprefix}}
\AtBeginBibliography{\togglefalse{blx@useprefix}}
\usepackage{tikz}
\usetikzlibrary{backgrounds,calc,chains,fit,matrix,positioning,shadows,shapes.misc,circuits.ee}
\tikzset{input/.style={}}
\tikzset{output/.style={}}
\tikzset{operator/.style={circle, draw, thin, fill=black!8, minimum size=2.5ex, inner sep=0pt}}
\tikzset{filter/.style={rectangle, draw, thin, fill=black!8, minimum size=3.5ex, inner xsep=1.5ex}}
\tikzset{other/.style={rounded rectangle, thin, draw, fill=black!8, minimum size=3.5ex, inner xsep=1ex}}
\tikzset{branch/.style={circle, draw, thin, fill=black, minimum size=.5ex, inner sep=0pt}}
\tikzset{rv/.style={circle, draw, thick, fill=white, minimum size=2.75ex, inner sep=0pt}}
\tikzset{ob/.style={circle, draw, thick, fill=lightgray, minimum size=2.75ex, inner sep=0pt}}
\tikzset{pa/.style={circle, draw, thick, fill=black, minimum size=1ex, inner sep=0pt}}
\tikzset{>=direction ee}

\usepackage{pgfplots}
\pgfplotsset{compat=1.14}
\pgfplotsset{every axis/.append style={enlargelimits={abs=3pt},grid,axis lines=left}}
\pgfplotsset{every axis plot/.append style={thick,mark size=1.5pt,line join=bevel,mark options={solid}}}
\pgfplotsset{label style={font=\small}}
\pgfplotsset{tick label style={font=\footnotesize}}
\pgfplotsset{grid style={color=black!10}}
\pgfplotsset{legend style={draw=none,opacity=.85,font=\footnotesize,cells={anchor=west,opacity=1}}}
\pgfplotsset{every non boxed x axis/.style={xtick align=center,shorten <=-.5\pgflinewidth}}
\pgfplotsset{every non boxed y axis/.style={ytick align=center,shorten <=-.5\pgflinewidth}}
\pgfplotsset{every non boxed z axis/.style={ztick align=center,shorten <=-.5\pgflinewidth}}
\pgfplotsset{/pgf/number format/1000 sep={\,}}

\newcommand{\E}{\operatorname{\mathbb E}}

\newcommand{\T}{\top}

\allowdisplaybreaks[2]
\begin{document}
\title{Efficient Nonlinear Transforms \\ for Lossy Image Compression}

\author{\IEEEauthorblockN{Johannes Ballé}
\IEEEauthorblockA{Google\\
Mountain View, CA 94043, USA\\
jballe@google.com}}


\maketitle

% IEEE copyright notice for arXiv (not for official submission)
% http://ieeeauthorcenter.ieee.org/publish-with-ieee/author-education-resources/guidelines-and-policies/policy-posting-your-article/
\tikz[remember picture,overlay] \node[above left,xshift=-1ex,yshift=1ex] at (current page.south east) {\copyright{} 2018 IEEE. Accepted as a conference contribution to Picture Coding Symposium 2018};%
\begin{abstract}
We assess the performance of two techniques in the context of nonlinear transform coding with artificial neural networks, Sadam and GDN. Both techniques have been successfully used in state-of-the-art image compression methods, but their performance has not been individually assessed to this point. Together, the techniques stabilize the training procedure of nonlinear image transforms and increase their capacity to approximate the (unknown) rate--distortion optimal transform functions. Besides comparing their performance to established alternatives, we detail the implementation of both methods and provide open-source code along with the paper.
\end{abstract}


% \fi
\IEEEpeerreviewmaketitle

\section{Introduction}
The hallmark of artificial neural networks (ANNs) is that, given the right set of parameters, they can approximate arbitrary functions~\cite{LeLiPiSc93}, and it appears that finding sufficiently good parameters is no longer a problem for many applications. Efforts to utilize techniques from machine learning, such as ANNs and stochastic gradient descent (SGD), in the context of image compression have recently garnered substantial interest~\cite{ToOMHwViMi16,BaLaSi16a,BaLaSi17,ThShCuHu17,ToViJoHwMi17,RiBo17,AgMeTsCaTi17,BaMiSiHwJo18}. Even though these models have been developed from scratch, they have rapidly become competitive with modern conventional compression methods (e.g.~\cite{BaMiSiHwJo18}, published only  about two years after the earliest publications appeared~\cite{ToOMHwViMi16,BaLaSi16a}), whereas conventional methods such as HEVC~\cite{HEVC} are the culmination of decades of engineering efforts. If properly utilized, ANNs may continue to enable rapid development of new image compression models, with fewer constraints to hinder the engineering process than before.

One such constraint, in the context of transform coding, is linearity. Arguably, the most well-known (and well-used) transform is the discrete cosine transform (DCT). However, its closeness to optimality in a rate--distortion sense has only been established under the assumption that the transform be linear, and the data distribution be Gaussian~\cite{AhNaRa74}. Although the vast majority of contemporary image compression methods are made up of nonlinear extensions to improve performance under empirical (non-Gaussian) data distributions, they rely on linear transforms at their core.

By way of \emph{nonlinear transform coding}, ANN-based methods discard this constraint. The analysis transform is replaced with a generic parametric function $\bm y = g_a(\bm x; \bm \phi)$, implemented by a neural network, where $\bm x$ is the image vector and $\bm \phi$ a (potentially large) set of parameters, and, analogously, the synthesis transform with a function $\bm{\hat x} = g_s(\bm{\hat y}; \bm \theta)$, where $\bm{\hat y} = Q(\bm y)$ represents the quantized image representation, $\bm \theta$ another set of parameters, and $\bm{\hat x}$ the reconstructed image vector. The parameters of the functions are obtained by minimizing an empirical rate--distortion loss function $L$, taking expectations over the (unknown) image data distribution $p_{\bm x}$:
\begin{equation}
L(\bm \phi, \bm \theta, \bm \psi) = \E_{\bm x \sim p_{\bm x}} \bigl[-\log p_{\bm{\hat y}}\bigl(\bm{\hat y}; \bm \psi\bigr) + \lambda\,d(\bm x, \bm{\hat x})\bigr],
\end{equation}
where the left-hand rate term contains an entropy model $p_{\bm{\hat y}}$ with parameters $\bm \psi$ that are jointly optimized, the right-hand term represents expected distortion as measured by a distortion metric $d$, and $\lambda$ is the Lagrange multiplier, determining the trade-off between the rate and distortion terms. Note that invertibility of the transforms is not theoretically guaranteed, but the minimization of the loss ensures reconstruction fidelity. Because the image distribution is unknown, the expectation is typically replaced with an average over a number of training images. With a further approximation to relax the discrete quantizer transfer curve (which would make derivatives useless as descent directions), this loss function can be made suitable for minimization with SGD~\cite{BaLaSi16a}.

While ANNs are generally composite functions, alternating between linear and nonlinear components, the quality of the solutions they offer can vary depending on what optimization algorithm, what constraints on the linear components (such as convolutionality), and what nonlinearities are used. Exactly how these details affect performance are poorly understood, which makes it necessary to resort to empirical evaluation.

The rest of this paper is concerned with evaluating two techniques which have been used in several recent compression models~\cite{BaLaSi16a,BaLaSi17,BaMiSiHwJo18}, but whose efficacy has not been individually evaluated: first, an extension to Adam~\cite{KiBa15}, a popular variant of SGD, which we call \emph{spectral Adam} or \emph{Sadam}, detailed in the following section; second, a nonlinearity introduced in~\cite{BaLaSi16} called generalized divisive normalization (GDN), described in section~\ref{sec:gdn}, which improves efficiency of the transforms compared to other popular nonlinearities. Both techniques make use of reparameterization, i.e., performing descent not on the transform parameters, but on invertible functions of them. We compare their performance to other established techniques and provide a detailed explanation of how we implemented them.

\section{Gradient conditioning with Sadam}
Consider a generalized loss function $L$, which is computed as the sum of some function of a linear projection of data vectors $\bm x$,
\begin{equation}
\label{eq:layer}
L = \sum_{\bm x} l(\bm z) \text{ with } \bm z = \bm H \bm x,
\end{equation}
where $\bm H$ is a matrix consisting of filters $\bm h_i$, such that $\bm H = [ \bm h_0, \bm h_1, \dotsc ]^\T$. The filters are a subset of the parameters to be optimized. For any one of the filters $\bm h$, the update rule of gradient descent dictates subtraction of the gradient of the loss function with respect to $\bm h$, multiplied by a step size $\rho$:
\begin{equation}
\Delta \bm h = -\rho \, \frac {\partial L} {\partial \bm h} = -\rho \sum_{\bm x} \frac {\partial l} {\partial z} \, \bm x,
\end{equation}
where $z$ is the element of $\bm z$ that corresponds to that filter. The updates $\Delta \bm h$ consist of scalar mixtures of the data vectors $\bm x$ that are projected onto the corresponding filter, and hence inherit much of the covariance structure of the data. Consequently, if $\bm x$ is drawn from a set of natural images, which have a power spectrum roughly inversely proportional to spatial frequency~\cite{Fi87}, the effective step size is several orders of magnitude higher for low-frequency components of $\bm h$ than high-frequency components, leading to an inbalance in convergence speeds, or even stability problems.

A classic remedy to this problem is pre-whitening of the data~\cite{LeBo99}, i.e., replacing $\bm x$ with $\bm P \bm x$, where $\bm P$ is a predetermined whitening matrix. This suggests the update rule
\begin{equation}
\Delta \bm h = -\rho \sum_{\bm x} \frac {\partial l} {\partial z} \, \bm P \bm x,
\end{equation}
where $\bm P \bm x$ is now white, and thus the effective step size is equalized across frequencies. However, this only works when $\bm H$ is applied directly to the data, and ANNs typically consist of several layers of linear--nonlinear functions. One can hope that decorrelated inputs will lead to less correlated intermediates, but in general, the covariance structure at higher layers is unknown a priori, and may still be ill-conditioned.

An adaptive scheme to conditioning the optimization problem is provided by the Adam algorithm~\cite{KiBa15}, a popular variation on stochastic gradient descent with an update rule
\begin{equation}
\label{eq:adam}
\Delta \bm h = -\rho \, \bm C_h^{-\frac 1 2} \bm m_h,
\end{equation}
where $\bm m_h$ is a running average of the derivative $\frac {\partial l} {\partial z} \, \bm x$, and $\bm C_h$ a diagonal matrix representing a running estimate of its covariance. Since $\bm C_h$ is constrained to be diagonal, however, it cannot effectively represent the covariance structure of natural images. To solve this, we apply the algorithm not to $\bm h$, but to its real-input discrete Fourier transform (RDFT), by reparameterizing
\begin{equation}
\label{eq:reparam}
\bm h = \bm F^\T \bm g \text{ (and } \bm g = \bm F \bm h \text{).}
\end{equation}
To optimize $\bm g$, we now use the derivative
\begin{equation}
\label{eq:derivative_g}
\frac {\partial l} {\partial \bm g} =
\bm F \frac {\partial l} {\partial \bm h} =
\frac {\partial l} {\partial z} \, \bm F \bm x
\end{equation}
and apply the update rule in \eqref{eq:adam} to it. Because \eqref{eq:reparam} is linear, the effective update to $\bm h$ can be computed as:
\begin{equation}
\Delta \bm h =
\bm F^\T \bigl(-\rho \, \bm C_g^{-\frac 1 2} \bm m_g \bigr) =
-\rho \, \bm F^\T \bm C_g^{-\frac 1 2} \bm F \, \bm m_h,
\end{equation}
where $\bm m_g$ and $\bm C_g$ are running estimates for the derivative in \eqref{eq:derivative_g}. As before, $\Delta \bm h$ consists of $\bm m_h$, rescaled with the inverse square root of a covariance estimate. However, the covariance estimate $\bm F^\T \bm C_g \bm F$ is diagonal in the Fourier domain, rather than diagonal in terms of the filter coefficients. As long as $\bm x$ has a shift-invariance property, like virtually all spatiotemporal data (including images, videos, or audio), the Fourier basis $\bm F$ is guaranteed to be a good approximation to the eigenvectors of the true covariance structure of $\bm x$ (up to boundary effects).

This implies that the modified algorithm can model the true covariance structure of the derivatives in a near-optimal fashion, using nothing more than two Fourier transforms applied to the filter coefficients. We call this technique \emph{Sadam}, or spectral Adam. Like Adam, and unlike pre-whitening, it adapts to the data and can be used in online training settings, and it can be applied to filters on any layer, as long as the neural network is convolutional (which preserves the shift-invariance property of each layer's inputs).

\section{Local normalization with GDN}
\label{sec:gdn}%
Local normalization has been known to occur ubiquitously throughout biological sensory systems, including the human visual system~\cite{CaHe12}, and has been shown to aid in factorizing the probability distribution of natural images~\cite{MaLa10,Ly10}. The reduction of statistical dependencies is thought to be an essential ingredient of transform coding. As such, it is not surprising that local normalization, as implemented by generalized divisive normalization (GDN)~\cite{BaLaSi16}, has recently been successfully applied to image compression with nonlinear transforms.

GDN is typically applied to linear filter responses $\bm z = \bm H \bm x$, where $\bm x$ are image data vectors, or to linear filter responses inside a composite function such as an ANN. Its general form is defined as
\begin{equation}
y_i = \frac {z_i} {\bigl( \beta_i + \sum_j \gamma_{ij} \, |z_j|^{\alpha_{ij}} \bigr)^{\varepsilon_i}},
\end{equation}
where $\bm y$ represents the vector of normalized responses, and vectors $\bm \beta$, $\bm \varepsilon$ and matrices $\bm \alpha$, $\bm \gamma$ represent parameters of the transformation (all non-negative). In some contexts, the values of $\bm \alpha$ and $\bm \varepsilon$ are fixed to make the denominator resemble a weighted $\ell^2$-norm (with an extra additive constant):
\begin{equation}
\label{eq:gdn_l2}
y_i = \frac {z_i} {\sqrt{ \beta_i + \sum_j \gamma_{ij} \, z_j^2 }}.
\end{equation}
In the context of convolutional neural networks (CNNs), GDN is often also constrained to operate locally in space, such that the indexes $i$, $j$ run across responses of different filters, but not across different spatial positions in the tensor~$\bm z$. These modifications serve to simplify implementation while retaining most of the flexibility of the transformation.

Generally, these parametric forms are simple enough to implement for given values of the parameters. However, optimizing the parameters using gradient descent can pose some practical caveats, which we address below. The derivatives of the GDN function with respect to its parameters are given by:
\begin{align}
\label{eq:beta_gradient}
\frac {\partial y_i} {\partial \beta_m} &= \delta_{im} \, z_i \cdot \Bigl( \beta_i + \sum_j \gamma_{ij} \, z_j^2 \Bigr)^{-\frac 3 2}, \\
\frac {\partial y_i} {\partial \gamma_{mn}} &= \delta_{im} \, z_i \, z_n^2 \cdot \Bigl( \beta_i + \sum_j \gamma_{ij} \, z_j^2 \Bigr)^{-\frac 3 2},
\end{align}
where $\delta$ is the Kronecker symbol. It is easy to see that as $\bm \beta$ and $\bm \gamma$ approach zero, the derivatives tend to grow without bounds. In practice, this can lead the optimization problem to become ill-conditioned. An effective remedy, as in the section before, is to use reparameterization. For example, consider any element $\beta$ of $\bm \beta$. Parameterizing it as $\beta = \nu^2$ would yield derivatives
\begin{equation}
\frac {\partial L} {\partial \nu} =
\frac {\partial L} {\partial \beta} \, \frac {\partial \beta} {\partial \nu} =
2 \nu \, \frac {\partial L} {\partial \beta},
\end{equation}
a gradient descent step would yield an updated value of
\begin{equation}
\nu' = \nu - 2 \nu \rho \, \frac {\partial L} {\partial \beta},
\end{equation}
and the effective update on $\beta$ can be computed as
\begin{align}
\label{eq:beta_update}
\Delta \beta = (\nu')^2 - \beta
&= \nu^2 - 4 \nu^2 \rho \, \frac {\partial L} {\partial \beta} + 4 \nu^2 \rho^2 \Bigl(\frac {\partial L} {\partial \beta}\Bigr)^2 - \beta \notag \\
&= -4 \beta \rho \Bigl( 1 - \rho \frac {\partial L} {\partial \beta} \Bigr) \frac {\partial L} {\partial \beta}.
\end{align}
Considering that, typically, $\rho \ll 1$, the expression in parentheses can be neglected. The effective step size on $\beta$ is hence approximately $4 \beta \rho$, which decreases linearly as $\beta$ approaches zero, mitigating the growth of the derivatives in \eqref{eq:beta_gradient} (while not completely eliminating it).

Two problems remain with this solution: first, should any parameter $\beta$ happen to be identical to zero at any point in the optimization, the next update as given by \eqref{eq:beta_update} would be zero as well, regardless of the loss function. This implies that the parameters are at risk to get “stuck” at zero. Second, negative values of the parameters need to prevented. To solve both, we use the reparameterization
\begin{equation}
\label{eq:gdn_reparam}
\beta = \max\Bigl(\nu, \sqrt{\beta_\textrm{min} + \epsilon^2}\Bigr)^2 - \epsilon^2,
\end{equation}
where $\epsilon$ is a small constant, and $\beta_\textrm{min}$ is a desired minimum value for $\beta$. We have found $\epsilon = 2^{-18}$ to work well empirically. To prevent the denominator in \eqref{eq:gdn_l2} to become zero, we set $\beta_\textrm{min} = 10^{-6}$, and use the same elementwise reparameterization for $\bm \gamma$, with $\gamma_\textrm{min} = 0$.

The reparameterization~\eqref{eq:gdn_reparam} can be implemented either using projected gradient descent, i.e., by alternating between a descent step on $\nu$ with $\beta = \nu^2 - \epsilon^2$ and a projection step setting $\nu$ to $\max(\nu, \sqrt{\beta_\textrm{min} + \epsilon^2})$, or by performing gradient descent directly on \eqref{eq:gdn_reparam}. In the latter case, it can be helpful to replace the gradient of the maximum operation $m = \max(\nu, c)$ with respect to the loss function $L$:
\begin{equation}
\frac {\partial L} {\partial \nu} = \frac {\partial L} {\partial m} \, \frac {\partial m} {\partial \nu} \coloneqq
\begin{cases}
\frac {\partial L} {\partial m} & \nu \ge c \text{ or } \frac {\partial L} {\partial m} < 0 \\
0 & \text{otherwise.}
\end{cases}
\end{equation}
This propagates updates to $\nu$ even if it is smaller than $c=\sqrt{\beta_\textrm{min} + \epsilon^2}$ (at which point the precise value of $\nu$ is irrelevant, and $\beta = \beta_\textrm{min}$). However, it only does so if the update would move $\nu$ closer to $c$, as further decreasing it would be pointless.

\section{Experimental results}
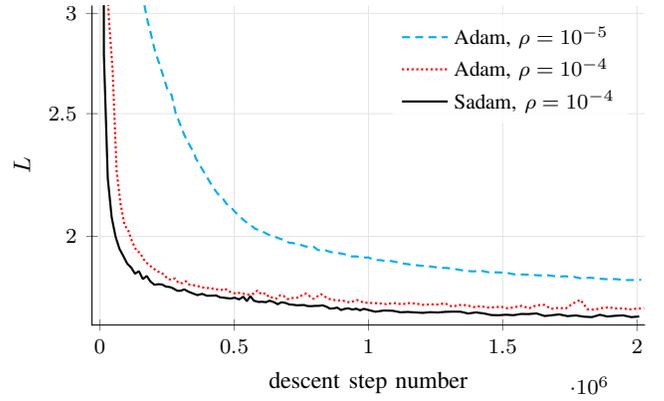
\begin{figure}
  \raggedleft
  \begin{tikzpicture}[trim axis right]
  \begin{axis}[
      height=.18\textheight,
      width=.83\linewidth,
      scale only axis,
      xlabel=descent step number,
      ylabel=$L$,
      ymode=log,
      log ticks with fixed point,
      ytick={2,2.5,3},
      ymax=3,
      xmin=0,
      xmax=2000000,
      legend pos=north east,
      ]
      \addplot[color=cyan,densely dashed] table[x=Step,y=Value,col sep=comma] {figures/adam5-128.csv};
      \addplot[color=red,densely dotted] table[x=Step,y=Value,col sep=comma] {figures/adam4-128.csv};
      \addplot[color=black] table[x=Step,y=Value,col sep=comma] {figures/sadam-128.csv};
      \legend{{Adam, $\rho=10^{-5}$},{Adam, $\rho=10^{-4}$},{Sadam, $\rho=10^{-4}$}};
  \end{axis}
  \end{tikzpicture}
  \caption{Rate--distortion performance over $2 \cdot 10^6$ descent steps on a compression model with the GDN nonlinearity and $N=128$ filters per layer. Sadam reduces the loss function faster than Adam with the same step size.}
  \label{fig:training128}
\end{figure}

\begin{figure}
  \raggedleft
  \begin{tikzpicture}[trim axis right]
  \begin{axis}[
      height=.18\textheight,
      width=.83\linewidth,
      scale only axis,
      xlabel=descent step number,
      ylabel=$L$,
      ymode=log,
      log ticks with fixed point,
      ytick={2,3,4,5,10},
      ymax=10,
      xmin=0,
      xmax=2000000,
      legend pos=north east,
      ]
      \addplot[color=cyan,densely dashed] table[x=Step,y=Value,col sep=comma] {figures/adam5-192.csv};
      \addplot[color=red,densely dotted] table[x=Step,y=Value,col sep=comma] {figures/adam4-192.csv};
      \addplot[color=black] table[x=Step,y=Value,col sep=comma] {figures/sadam-192.csv};
      \legend{{Adam, $\rho=10^{-5}$},{Adam, $\rho=10^{-4}$},{Sadam, $\rho=10^{-4}$}};
  \end{axis}
  \end{tikzpicture}
  \caption{Rate--distortion performance over $2 \cdot 10^6$ descent steps on a compression model with the GDN nonlinearity and $N=192$ filters per layer. For this setup, the step size $\rho=10^{-4}$ is too large for Adam, leading to instabilities which can cause unpredictable outcomes and make experimentation difficult.}
  \label{fig:training192}
\end{figure}
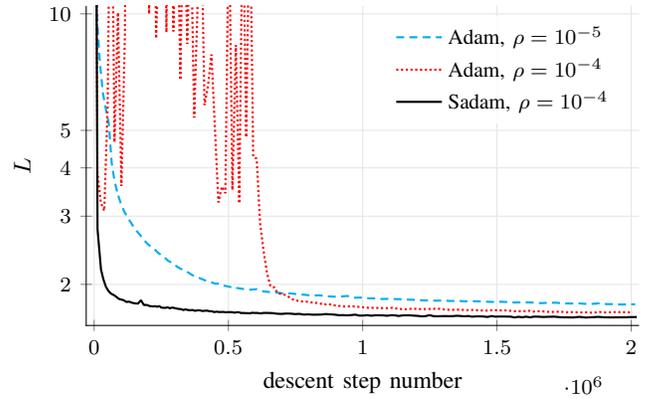

\begin{figure}
  \raggedleft
  \begin{tikzpicture}[trim axis right]
  \begin{axis}[
      height=.21\textheight,
      width=.85\linewidth,
      scale only axis,
      xlabel=bits per pixel,
      ylabel=PSNR,
      xmin=.1,
      xmax=1.5,
      ymin=27,
      legend pos=south east,
      reverse legend,
      ]
      \addplot[color=cyan,densely dashed,mark=*] table {
        %tigger-pcs2018-adam0.00001-128-exp6 [/ vs. /]:
        0.121674 26.294566
        0.306256 29.044121
        0.740871 32.647087
        1.267145 34.397452
      };
      \addplot[color=cyan,densely dashed,mark=*] table {
        %tigger-pcs2018-adam0.00001-192-exp6 [/ vs. /]:
        0.119624 26.315916
        0.315131 29.398211
        0.745517 32.933783
        1.506011 36.692710
      };
      \addplot[color=red,densely dotted,mark=*] table {
        %tigger-pcs2018-adam0.0001-128-exp6 [/ vs. /]:
        0.116971 26.416929
        0.300164 29.363839
        0.708888 32.856968
        1.204882 34.741857
      };
      \addplot[color=red,densely dotted,mark=*] table {
        %tigger-pcs2018-adam0.0001-192-exp6 [/ vs. /]:
        0.119954 26.523468
        0.296057 29.353084
        0.717828 33.127562
        1.439444 36.945758
      };
      \addplot[color=black,mark=*] table {
        %tigger-pcs2018-baseline-128-exp6 [/ vs. /]:
        0.122107 26.563073
        0.312878 29.552105
        0.724862 33.045169
        1.204744 34.844298
      };
      \addplot[color=black,mark=*] table {
        %tigger-pcs2018-baseline-192-exp6 [/ vs. /]:
        0.121719 26.669928
        0.316634 29.699740
        0.739410 33.466862
        1.415743 36.960420
      };
      \legend{{Adam, $\rho=10^{-5}$},,{Adam, $\rho=10^{-4}$},,{Sadam, $\rho=10^{-4}$},};
      \path (axis cs:0.740871,32.647087) -- (axis cs:1.267145,34.397452) node[midway,sloped,below] {\footnotesize $N=128$};
      \path (axis cs:0.739410,33.466862) -- (axis cs:1.415743,36.960420) node[midway,sloped,above] {\footnotesize $N=192$};
  \end{axis}
  \end{tikzpicture}
  \caption{Rate--distortion performance after $2 \cdot 10^6$ descent steps with different training algorithms. All algorithms converge to similar results eventually, but Sadam remains better when comparing the same number of steps.}
  \label{fig:sadam}
\end{figure}
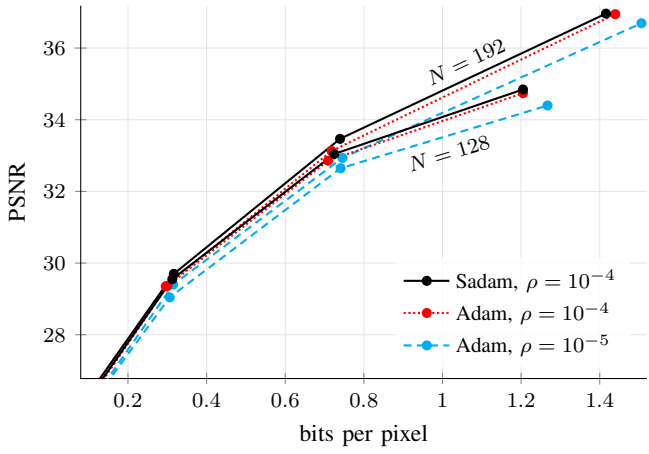

\begin{figure}
  \raggedleft
  \begin{tikzpicture}[trim axis right]
  \begin{axis}[
      height=.21\textheight,
      width=.85\linewidth,
      scale only axis,
      xlabel=bits per pixel,
      ylabel=PSNR,
      xmin=.1,
      xmax=1.5,
      ymin=27,
      legend pos=south east,
      reverse legend,
      ]
      \addplot[color=teal,densely dash dot dot,mark=*] table {
        %tigger-pcs2018-none-128-exp6 [/ vs. /]:
        0.155632 26.033283
        0.442378 29.421533
        0.943214 32.823605
        1.354017 33.660986
      };
      \addplot[color=cyan,mark=*] table {
        %tigger-pcs2018-tanh-128-exp6 [/ vs. /]:
        0.115200 26.176700
        0.295969 28.937674
        0.701259 32.324823
        1.150096 33.832727
      };
      \addplot[color=blue,densely dash dot,mark=*] table {
        %tigger-pcs2018-softplus-128-exp6 [/ vs. /]:
        0.111376 26.095697
        0.281281 28.711208
        0.669230 31.978703
        1.202403 34.142880
      };
      \addplot[color=purple,densely dotted,mark=*] table {
        %tigger-pcs2018-relu-128-exp6 [/ vs. /]:
        0.115058 26.263206
        0.290678 28.845951
        0.723275 32.423437
        1.205378 33.931541
      };
      \addplot[color=red,densely dashed,mark=*] table {
        %tigger-pcs2018-leaky_relu-128-exp6 [/ vs. /]:
        0.116096 26.273533
        0.291679 28.979856
        0.719771 32.377154
        1.204885 34.168158
      };
      \addplot[color=black,mark=*] table {
        %tigger-pcs2018-baseline-128-exp6 [/ vs. /]:
        0.122107 26.563073
        0.312878 29.552105
        0.724862 33.045169
        1.204744 34.844298
      };
      \legend{none (linear),tanh,softplus,ReLU,leaky ReLU,GDN};
  \end{axis}
  \end{tikzpicture}
  \caption{Rate--distortion performance after $2 \cdot 10^6$ descent steps with different nonlinearities in a CNN with $N=128$ filters per layer. GDN increases the approximation capacity of the network, leading to better performance.}
  \label{fig:nonlinearity128}
\end{figure}
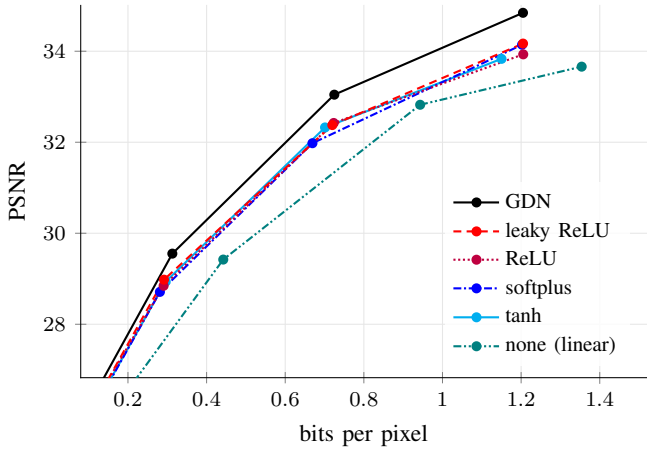

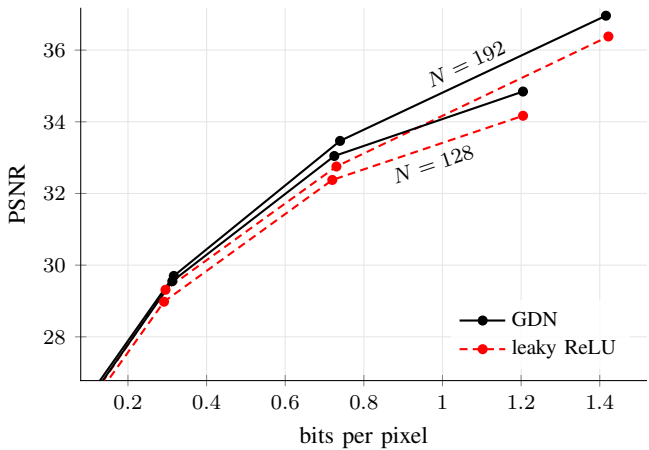
\begin{figure}
  \raggedleft
  \begin{tikzpicture}[trim axis right]
  \begin{axis}[
      height=.21\textheight,
      width=.85\linewidth,
      scale only axis,
      xlabel=bits per pixel,
      ylabel=PSNR,
      xmin=.1,
      xmax=1.5,
      ymin=27,
      legend pos=south east,
      reverse legend,
      ]
      \addplot[color=red,densely dashed,mark=*] table {
        %tigger-pcs2018-leaky_relu-128-exp6 [/ vs. /]:
        0.116096 26.273533
        0.291679 28.979856
        0.719771 32.377154
        1.204885 34.168158
      };
      \addplot[color=red,densely dashed,mark=*] table {
        %tigger-pcs2018-leaky_relu-192-exp6 [/ vs. /]:
        0.115896 26.429854
        0.295463 29.312575
        0.730248 32.752280
        1.421960 36.379937
      };
      \addplot[color=black,mark=*] table {
        %tigger-pcs2018-baseline-128-exp6 [/ vs. /]:
        0.122107 26.563073
        0.312878 29.552105
        0.724862 33.045169
        1.204744 34.844298
      };
      \addplot[color=black,mark=*] table {
        %tigger-pcs2018-baseline-192-exp6 [/ vs. /]:
        0.121719 26.669928
        0.316634 29.699740
        0.739410 33.466862
        1.415743 36.960420
      };
      \legend{leaky ReLU,,GDN,};
      \path (axis cs:0.719771,32.377154) -- (axis cs:1.204885,34.168158) node[midway,sloped,below] {\footnotesize $N=128$};
      \path (axis cs:0.739410,33.466862) -- (axis cs:1.415743,36.960420) node[midway,sloped,above] {\footnotesize $N=192$};
  \end{axis}
  \end{tikzpicture}
  \caption{Rate--distortion performance after $2 \cdot 10^6$ descent steps with different nonlinearities, comparing networks with different number of filters $N$. Higher bitrates require a greater approximation capacity, making differences between nonlinearities more visible.}
  \label{fig:capacity}
\end{figure}

To evaluate the performance of the two methods introduced in the previous sections, we conducted experiments in the TensorFlow framework, using compression models very similar to the ones in~\cite{BaLaSi17} as a baseline for comparison.

The transforms were constructed as follows (refer to~\cite{BaLaSi17} for details). The analysis transform consists of a CNN with three convolutional layers, each with $N$ filters, a bias term (additive constant) for each filter, and zero boundary handling. We conducted two sets of experiments, where we set $N=128$ or $N=192$, respectively. The kernel support is $9\times 9$, $5\times 5$, and $5\times 5$, and the downsampling factors $4\times 4$, $2\times 2$, and $2\times 2$, for each layer, in order. Analogously, the synthesis transform is a CNN with three layers with $N$ filters each (except for the last, which has 3 filters to match the three channels of RGB images), with kernel support and upsampling factors mirroring the analysis transform (i.e., $5\times 5$, $5\times 5$, and $9\times 9$, as well as $2\times 2$, $2\times 2$, and $4\times 4$, in order). Sandwiched between the convolutional layers are two nonlinear layers -- GDN or other popular nonlinearities for comparison: ReLU, leaky ReLU, softplus, tanh, and no nonlinearity, which makes the entire network linear. For the experiment involving GDN, we used inverse GDN (replacing division with multiplication) for the synthesis transform, as in~\cite{BaLaSi17}. For the entropy model $p_{\hat y}$, we used a separate non-parametric scalar density model for each filter, described in more detail in \cite[appendix 6.1]{BaMiSiHwJo18}.

We trained the model on random patches of $128\times 128$ pixels taken from a large dataset of web images, with preprocessing applied as described in \cite{BaMiSiHwJo18}. To avoid overfitting issues, we evaluated compression performance by averaging over the well-known Kodak dataset. All reported bit rates are average bitstream lengths, not cross entropy measurements.

The first three figures compare training algorithms. Generally, the convergence behavior can differ depending on the other hyperparameters, such as the number of filters~$N$ (figures~\ref{fig:training128} and~\ref{fig:training192}), but Sadam outperforms Adam in all experiments when comparing an identical number of descent steps (figure~\ref{fig:sadam}). Note that a step size of $\rho = 10^{-4}$ in some setups appears too large for Adam, which can result in the optimization being unstable (figure~\ref{fig:training192}). Lowering the step size to $10^{-5}$ fixes this problem, but slows down convergence. Sadam achieves faster convergence in all cases, because it improves the conditioning of the optimization problem, allowing the same step size to be used across a variety of setups.

In the next two figures, we compare models which were all trained with Sadam and differ only in their nonlinearities (note that Sadam only affects the linear filter coefficients, not the GDN parameters). GDN is better than other nonlinearities at approximating the rate--distortion optimal transforms across all values of $\lambda$ and for different network architectures (figure~\ref{fig:nonlinearity128} only shows $N=128$; results for $N=192$ are similar), with an improvement of approximately 0.6--0.7dB PSNR at the highest rate point. Note that all models have approximately the same number of parameters for a given $N$ (GDN increases the number of parameters by less than 2\%; the other nonlinearities have no parameters). Computationally, the convolutions tend to eclipse the nonlinearities (including GDN).

We observe that differences between nonlinearities manifest themselves mostly at the higher bitrates (for large values of~$\lambda$). Moreover, many final-layer filters of the analysis transforms end up not being utilized for lower bitrate models, converging to produce constant outputs with zero entropy (not shown). This can be interpreted in the context of \emph{approximation capacity}. A higher number of filters $N$ increases the network's degrees of freedom, and hence its capacity to approximate the rate--distortion optimal transform (figure~\ref{fig:capacity}). Because we use the same $N$ across different values of $\lambda$, and because a higher bitrate (i.e., information bandwidth) implies that the optimal set of transforms is more complex, or higher-dimensional, the approximation capacity of the models is driven to its limit there. At lower bitrates, where the approximation capacity of all models is sufficient, their rate--distortion performance converges, making them simply different approximations of an equivalent set of transforms. Note that the GDN network with $N=128$ achieves similar performance to the leaky ReLU network with $N=192$ at approximately 0.9 bits/pixel. GDN may be able to approximate the optimal image transforms using fewer filters than other nonlinearities, because it has a parametric form suitable for redundancy reduction in images.

Note that we experimented with batch normalization~\cite{IoSz15} as well, but found that it lead to substantially worse results in the given setup, regardless of the nonlinearity used, even when increasing the batch size from 16 to values as high as 128, which is close to the feasible limit for training on a single GPU. This may be due to a higher tolerance for internal noise in ANNs trained for classification as opposed to compression.

\section{Conclusion}
We present two techniques, Sadam and GDN, to improve the training and representation efficiency of ANN-based image transforms in the context of nonlinear transform coding, and provide a detailed comparison of their performance with other popular training algorithms and nonlinearities. Sadam stabilizes and speeds up the training procedure for all experiments. GDN appears to increase the approximation capacity of the transforms, suggesting that networks with fewer filters can be used with GDN, compared to other nonlinearities, without hurting performance. We provide implementations of Sadam, GDN, and the entropy model at \url{https://github.com/tensorflow/compression}. Given the encouraging results, a topic for future work is to explore whether the increased efficiency of GDN in terms of number of filters also translates into a computational efficiency advantage, depending on the hardware platform it is implemented on, and to further explore the relationship between batch normalization and GDN.

\printbibliography

\end{document}